\documentclass[runningheads]{llncs}

\usepackage{amsfonts}
\usepackage{amsmath}
\usepackage{amssymb}
\usepackage{mathtools}
\usepackage{color}

\usepackage{eqparbox}

\usepackage{graphicx}
\usepackage{multirow}
\usepackage{pbox}
\usepackage{float}
\usepackage{caption}
\usepackage{subcaption}

\usepackage{pgfplots}
\usepackage{tikz}
\usetikzlibrary{bayesnet}
\usetikzlibrary{arrows}
\usetikzlibrary{backgrounds}
\usetikzlibrary{calc}

\DeclarePairedDelimiterX{\infdivx}[2]{(}{)}{#1\;\delimsize\|\;#2}
\newcommand{\KL}{\text{KL}\infdivx}
\newcommand{\xoverbrace}[2][\vphantom{\dfrac{A}{A}}]{\overbrace{#1#2}}

\newcommand{\x}[0]{\mathbf{x}}
\newcommand{\y}[0]{\mathbf{y}}
\newcommand{\z}[0]{\mathbf{z}}
\newcommand{\SSigma}[0]{\mathbf{\Sigma}}
\newcommand{\mmu}[0]{\boldsymbol{\mu}}
\newcommand{\LLambda}[0]{\mathbf{\Lambda}}
\newcommand{\E}{\mathop{\mathbb{E}}\nolimits}

\begin{document}

\title{Deep Group-wise Variational \\
Diffeomorphic Image Registration}
\author{
Tycho F.A. van der Ouderaa \inst{1} \and
Ivana Išgum \inst{2,3} \and \\
Wouter B. Veldhuis \inst{4} \and
Bob D. de Vos \inst{1}
}

\authorrunning{Tycho F.A. van der Ouderaa, et al.}

\institute{
Quantib-U, Utrecht, The Netherlands \\
\small Correspondence to: \email{t.vanderouderaa@quantib.com} \and
Department of Biomedical Engineering and Physics, Amsterdam University Medical Center, University of Amsterdam, 
\and
Department of Radiology and Nuclear Medicine, Amsterdam University Medical Center, University of Amsterdam, Amsterdam, 
\and
Department of Radiology, University Medical Center Utrecht, Utrecht
}

\maketitle 

\begin{abstract}
Deep neural networks are increasingly used for pair-wise image registration. We propose to extend current learning-based image registration to allow simultaneous registration of multiple images. To achieve this, we build upon the pair-wise variational and diffeomorphic VoxelMorph approach and present a general mathematical framework that enables both registration of multiple images to their geodesic average and registration in which any of the available images can be used as a fixed image. In addition, we provide a likelihood based on normalized mutual information, a well-known image similarity metric in registration, between multiple images, and a prior that allows for explicit control over the viscous fluid energy to effectively regularize deformations. We trained and evaluated our approach using intra-patient registration of breast MRI and Thoracic 4DCT exams acquired over multiple time points. Comparison with Elastix and VoxelMorph demonstrates competitive quantitative performance of the proposed method in terms of image similarity and reference landmark distances at significantly faster registration.

\end{abstract}

\section{Introduction}

Image registration, the process of aligning images, is a fundamental problem in medical image analysis \cite{viergever2016survey}. For example, it can be used to align image volumes in longitudinal studies \cite{reuter2011avoiding}, or to align images acquired temporally during contrast enhancement \cite{rueckert1998non}.

A common way to register images is to apply a transformation, such as an affine transformation, B-splines \cite{rueckert1999nonrigid}, or a deformation (vector) field \cite{sotiras2013deformable}, and to optimize its parameters using some form of numerical optimization. The objective is to minimize misalignment defined by a loss, such as a squared error, negative normalized cross correlation, or more advanced dissimilarity measures such as negative normalized mutual information \cite{pluim2003mutual}. Additionally, unrealistic deformations are mitigated by regularization or by using a constraint family of deformations.

Methods that constrain deformation fields to diffeomorphisms, such as LDDMM \cite{beg2005computing}, Diffeomorphic Demons \cite{vercauteren2009diffeomorphic}, and SyN \cite{avants2008symmetric}, guarantee preservation of topology, and have shown to be very effective.

Recently, it has been shown that deep learning models can be trained to perform image registration, leveraging on the increased availability of medical imaging data \cite{de2019deep,balakrishnan2018unsupervised}. Once a model is trained, deformation fields can be obtained by inference, which is much faster than aligning images through an iterative optimization process. In \cite{dalca2018unsupervised} and \cite{dalca2019unsupervised}, diffeomorphic neural network-based methods were proposed, further bridging the gap between learning-based and classical approaches. Although deep learning methods show great potential, they are restricted to image pairs or require a heuristically constructed template \cite{che2019deep}.

In this work, we extend learning-based methods to enable fast group-wise image registration. We build upon the variational framework proposed by \cite{dalca2019unsupervised} and stack multiple images along the channel axis of the input and output of the network. In addition, we propose a likelihood based on normalized mutual information and a prior which yields explicit control over the \textit{viscous fluid energy}. We quantitatively evaluate our model and compare it with Elastix \cite{klein2009elastix} and Voxelmorph \cite{balakrishnan2019voxelmorph}. 

\section{Method}
\noindent
Let us consider random variables $\x$ and $\y$ each consisting of a collection of image volumes acquired at $K$ different timepoints. We consider $\y$ to be unknown aligned ground-truth and let $\z$ be a latent variable that parameterizes transformations $\phi_\z^{-1}$ that describes the formation of $\x$ as a transformation of $\y$ through $\phi_\z^{-1} \circ \y$. All variables can be partitioned into different subsets for each timepoint $k \in \{1,...,K\}$, where $\z_k \in \z$ defines a transformation on 3D voxels $\phi_{\z_k}^{\small-1}: \mathbb{R}^3 \to \mathbb{R}^3$ that maps the volume $\y_k \in \y$ onto volume $\x_k \in \x$ via $\phi_{\z_k}^{\small-1} \circ \y_k$. 

Depending on the task, we define images that require registration (i.e. moving images) as observations of $\x$ and the target images (i.e. fixed images) as observations of $\y$. We are not restricted to a predefined fixed timepoint and can describe all of our data as observations of $\x$, effectively letting $\y$ be an unobserved latent variable. We will refer to this case as \textit{all-moving} group-wise image registration, and the case where one timepoint is chosen as fixed image as \textit{all-to-one} group-wise image registration. Observe that we can obtain the former from the latter by constraining the deformation $\phi_{\z_p}^{-1}=\text{Id}$ at a fixed timepoint $p$ to be the identity. Moreover, we can view pair-wise registration as a particular $K=2$ instance in our group-wise framework, where one timepoint is kept fixed and the other one can move.

\subsection{Generative Model}
We propose a generative model and aim to find the posterior probability of our transformation parameters $p(\mathbf{z}|\x, \mathbf{y})$ that govern the deformations $\mathbf{\phi_z}$. Because the posterior is intractable we approximate it using \textit{variational inference}.

\subsubsection{Prior}
We define a multivariate normal prior $\mathbf{z}$ with zero mean and precision matrix $\LLambda$ as the inverse of the covariance:
\begin{equation}
p(\mathbf{z}) = \mathcal{N}(\mathbf{z} \mid 0, \LLambda^{-1}),
\end{equation}
where we choose $\LLambda=\lambda_v \mathbf{I} + \lambda_u \mathbf{L}$ with identity matrix $\mathbf{I}$ weighted by $\lambda_v$ and the Laplacian neighbourhood graph $\mathbf{L}=\mathbf{D}-\mathbf{A}$ with degree matrix $\mathbf{\mathbf{D}}$ and adjacency matrix $\mathbf{A}$, weighted by $\lambda_u$. We treat $\lambda_v$ and $\lambda_u$ as hyper-parameters that give control over the norm on the vectors in the field parameterized by $\z$ and the norm on its displacement, respectively. The prior is inspired on the spatial smoothing prior proposed in \cite{dalca2018unsupervised}, which can be seen as a particular $\lambda_v=0$ instance of our prior. If $\z$ describes the deformation field as is done in \cite{balakrishnan2019voxelmorph}, it penalizes the \textit{linear elastic energy} penalty \cite{miller1993mathematical} and in case $\phi$ is diffeomorphic and $\z$ describes the velocity of the deformation field, the prior is more akin to a fluid-based model and penalizes the \textit{viscous fluid energy} \cite{christensen1994deformable,bro1996fast,tian2013molecular}.

\begin{figure}[t]
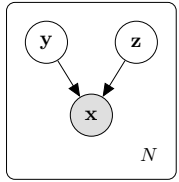

\medskip
\centering

\begin{subfigure}[b]{0.40\textwidth}
\centering
\resizebox{0.50\linewidth}{!}{
    \tikz{
        \node[obs] (x) {$\x$};
        \node[latent, above=of x, yshift=-0.5cm, xshift=-0.75cm] (y) {$\y$};
        \node[latent, above=of x, yshift=-0.5cm, xshift=0.75cm] (z) {$\z$};
        
       

        \plate [inner sep=.30cm] {plate_outer} {(z)(x)(y)} {$N$};
        
        \edge {y,z} {x}
    } 
}
\label{fig:1}
\end{subfigure}

\caption{Graphical model describing misaligned observations $\x$ as noisy observations of unknown aligned images $\y$ that have been transformed via $\phi_\z^{-1} \circ \y$.}
\label{fig:graphical_models}
\end{figure}

\subsubsection{Likelihood}
It is well known that (normalized) mutual information information is an effective similarity metric for image registration which is robust to varying intensity inhomogeneities \cite{viola1997alignment,maes1997multimodality}. Therefore, we model the probability of $\x$ as an exponential distribution of the average normalized mutual information (NMI) across all timepoint permutations:

\begin{equation}
p(\x|\y,\z)
\propto \exp \Big\{
\frac{1}{K(K-1)}\sum_{i\neq j} \text{NMI}(\phi_{\z_i} \circ \x_i, \phi_{\z_j} \circ \x_j)
\Big\}
\end{equation}
where we define $\phi_{\z_p}= \text{Id}$ and $\x_p = \y_p$ in case of a fixed timepoint $p$. Various approaches exist to estimate the probability density functions (both marginal and joint) of the intensity values needed to compute the entropies for the mutual information \cite{pluim2003mutual}. In this work, we obtain differentiable estimates of the intensity values using a Gaussian Parzen window estimation \cite{collignon1995automated,maes1997multimodality} of the voxel intensities in each mini batch.

\begin{figure}[t]
    \centering
    \resizebox{\linewidth}{!}{
        \includegraphics[]{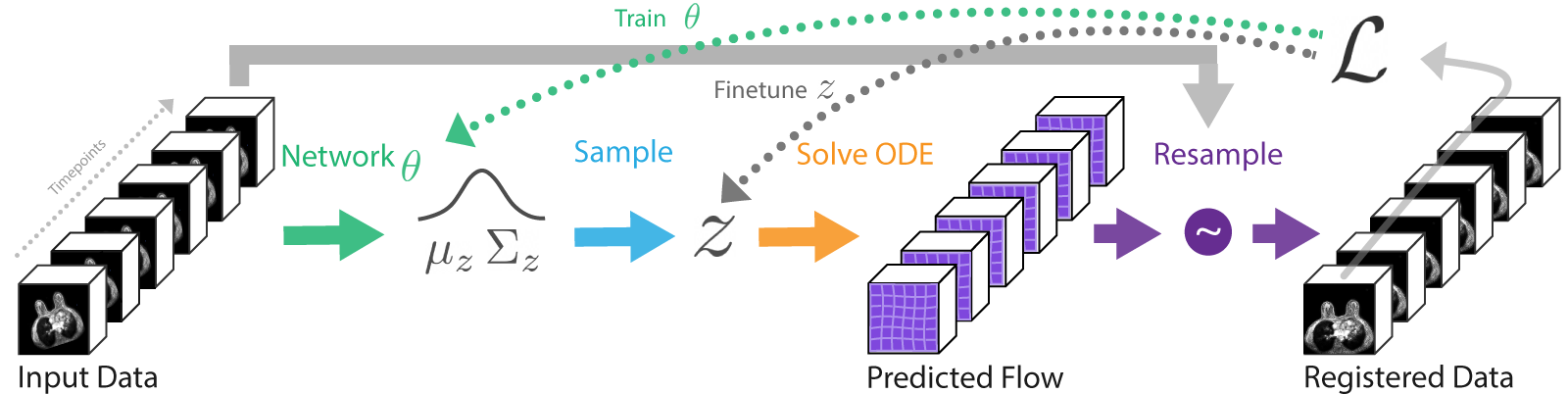}
    }
    \caption{GroupMorph: A neural network parameterized by $\theta$ is trained to output estimates over parameters of velocity fields $\z$ that can be integrated to obtain diffeomorphic flows that align the input images according to some loss $\mathcal{L}$.} \label{fig:training_procedure}
\end{figure}

\subsection{Variational Lower Bound}

Since the posterior probability $p(\z|\x,\y)$ is intractable, we cannot directly obtain MAP estimates of $\z$. Instead, we estimate the true posterior with a variational approximating posterior by minimizing the negative evidence lower bound $\mathcal{L}(\theta|\x,\y)$ which can be derived using the fact that the $\text{KL}$-divergence is non-negative \cite{kingma2013auto}:

\begin{align}
\log p_\theta(\x,\y) &= \KL{q_\theta(\z|\x,\y)}{ p(\z|\x,\y)} - \mathcal{L}(\theta|\x,\y) \\
\geq -\mathcal{L}(\theta|\x,\y) &= -\E_q \big[ -\log q_\theta(\z|\x,\y) + \log p(\x,\y,\z) \big] \\
& = -\E_q \left[ \log p(\x|\y,\z)\right] + \KL{q_\theta (\z|\x,\y)}{ p(\z)} + \text{const.}
\end{align}


\noindent
We let our variational approximate posterior \cite{kingma2013auto} be a multivariate Gaussian:
\begin{equation}
\log q_\theta(\z|\x,\y) = \log \mathcal{N}(\z | \mmu_\z, \SSigma_\z),
\end{equation} 
where mean $\mmu_\z$ and covariance matrix $\SSigma_\z$ of the approximate posterior $q_\theta(\z|\x,\y)$ are outputted by a neural network with parameters $\theta$. To avoid differentiating through a random variable, we obtain the sample $\z$ using the \textit{re-parameterization trick} \cite{kingma2013auto} $\z = \mmu_\z + \SSigma_\z \SSigma_\z^T e$ where $e \sim \mathcal{N}(0, \mathbf{I})$ is normally distributed. 

\subsection{Resulting Loss}
We find the resulting loss for a single Monte Carlo sample from $q$:

\begin{align}
\label{eq:final_loss}
\mathcal{L}(\theta|\x,\y) &=
\xoverbrace{-\E_q [ \log p(\x|\y,\z) ]}^{\text{\small Reconstruction Error}} + \xoverbrace{\text{KL}[ q_\theta(\z|\x,\y)||p(\z)]}^{\text{\small Regularization}} \\
&\simeq
\frac{1}{K(K-1)}\sum_{i\neq j} -\text{NMI}(\phi_{\z_i} \circ \x_i, \phi_{\z_j} \circ \x_j)
+
\frac{1}{2} \Big[ \text{tr}((\lambda_u \mathbf{D} + \lambda_v)\SSigma_\z - \log \SSigma_\z) + \mmu_\z^T \LLambda^{-1} \mmu_\z \Big] + \text{const.} \nonumber
\end{align}
where we define $\phi_{\z_p}= \text{Id}$ and $\x_p = \y_p$ in case of a fixed timepoint $p$ and where we can expand the last term $\mmu_\z^T \LLambda^{-1} \mmu_\z = \lambda_u \frac{1}{2} \sum_i\sum_{j\in N(i)} (\mmu[i] - \mmu[j])^2 + \lambda_v \sum_i \mmu[i]^2$ where $N(i)$ are the neighbors of voxel $i$.

\subsection{Diffeomorphic Transformations}

Diffeomorphic deformation fields are invertible ($\phi^{-1}$ exists) and have found to be very valuable in image registration as they preserve topology, by design. For diffeomorphic GroupMorph, we follow \cite{balakrishnan2019voxelmorph} and describe transformations $\phi_{\z_k}$ for each timepoints $k$ as an ordinary differential equation (ODE):

\begin{equation}
\frac{\partial \phi_{\z_k}^{(t)}}{\partial t} = v_{\z_k}(\phi_{\z_k}^{(t)})
\end{equation}
where $\phi_{\z_k}^{(0)}=\text{Id}$ is the identity transformation and $t$ is time. For every timepoint $k$, we let $\z_k$ parameterize the vectors of the velocity field $v_{\z_k}$ and solve deformation field $\phi_{\z_k}^{(1)}$ using the \textit{scaling and squaring} solver, proposed by \cite{dalca2018unsupervised} in the context of neural-based diffeomorphic image registration. For non-diffeomorphic models, we let $\z_k$ parameterize the vectors of the deformation field $\phi_{\z_k}$ directly.

\section{Implementation}

\begin{figure}[t]
    \centering
    \resizebox{\linewidth}{!}{
        \includegraphics[]{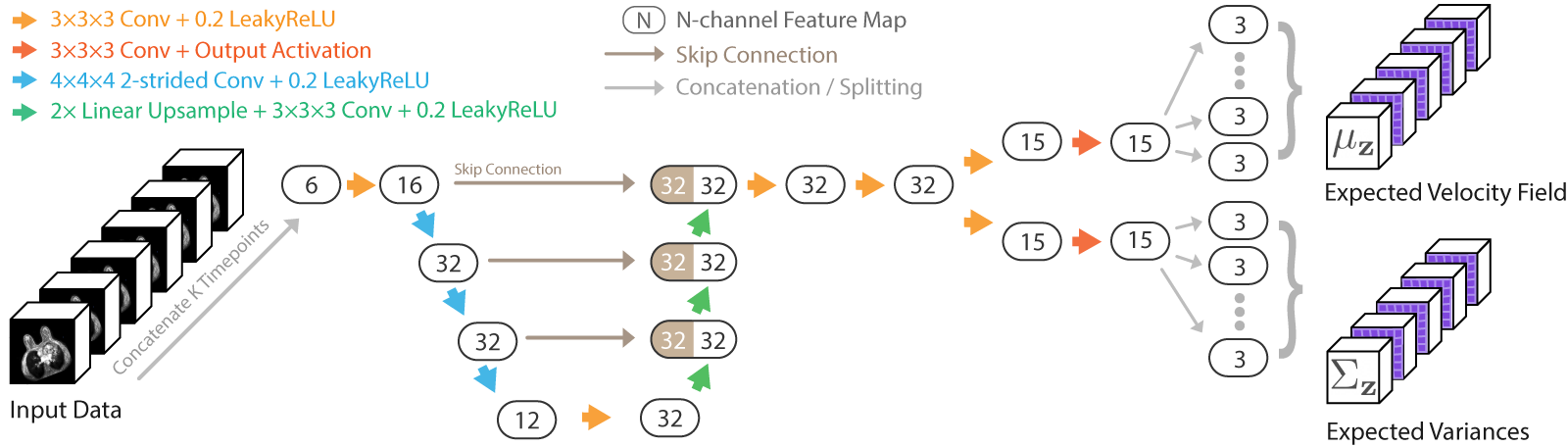}
    }
    \caption{Illustration of the 3D U-Net architecture used to output mean $\mu_\z$ and variance $\Sigma_\z$ of distribution over velocity fields $q_\theta(\z|\x,\y)$ used for diffeomorphic image registration.}
    \label{fig:network_architecture}
\end{figure}

We let a 3D U-net \cite{ronneberger2015u,balakrishnan2018unsupervised,balakrishnan2019voxelmorph}, output the parameters $\mmu_\z$ and $\SSigma_\z$ of the approximate posterior $q_\theta(\z|\x,\y)$, as illustrated in Figure \ref{fig:training_procedure}. Fixed and moving image volumes are concatenated along the channel dimension and form the input of the network. Figure \ref{fig:network_architecture} gives a more detailed description of the network architecture, including filter sizes and channel counts. All layers are followed by a Leaky ReLU \cite{maas2013rectifier} with 0.2 slope, except for the $\mmu_\z$ and $\SSigma_\z$ heads which are respectively followed by a hyperbolic tangent scaled with $\alpha=15$ that limits the output to $[-\alpha, \alpha]$ voxels and a Softplus \footnote{$\text{Softplus}(x)=\log(\exp(x)+1)$} activation to ensure positive variances.

We train for 100k iterations using Adam \cite{kingma2014adam} with a batch size of 1 and a learning rate of $1e^{-5}$ ($\beta_1=0.9, \beta_2=0.999$) decayed with cosine annealing \cite{loshchilov2016sgdr}. Inputs are normalized using 1-99 percentile normalization \cite{patrice2018image} and training samples consist of $96 \times 96 \times 96$ patches for all timepoints at a randomly sampled location. For the hyperparameters we choose $\lambda_u=1$ and $\lambda_v=0.01$.


\newpage
\section{Datasets}

\begin{figure}[t]
\resizebox{1.0\linewidth}{!}{
\begin{tabular}{c | c c c | c c c c }
No Registration & Elastix & VoxelMorph & VoxelMorph-diff & GroupMorph \textit{all-to-one} \textbf{(ours)} & GroupMorph-diff \textit{all-to-one} \textbf{(ours)} & GroupMorph \textit{all-moving} \textbf{(ours)} & GroupMorph-diff \textit{all-moving} \textbf{(ours)}  \\
\includegraphics[height=3.9cm,width=3.9cm]{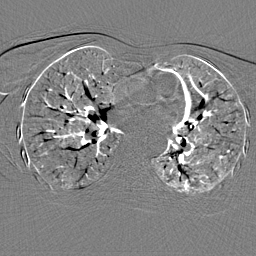} &
\includegraphics[height=3.9cm,width=3.9cm]{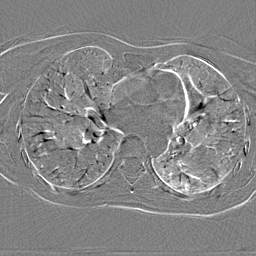} &
\includegraphics[height=3.9cm,width=3.9cm]{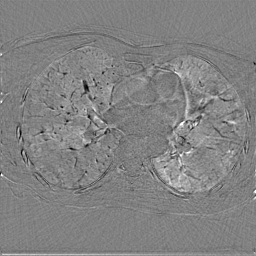} &
\includegraphics[height=3.9cm,width=3.9cm]{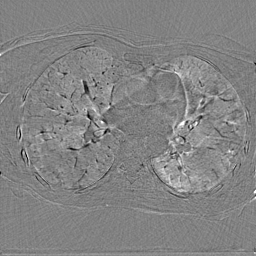} &
\includegraphics[height=3.9cm,width=3.9cm]{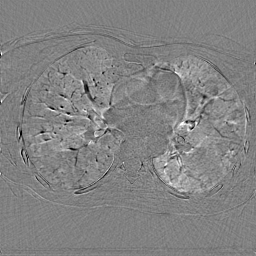} &
\includegraphics[height=3.9cm,width=3.9cm]{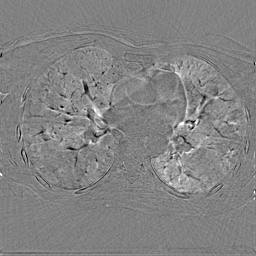} &
\includegraphics[height=3.9cm,width=3.9cm]{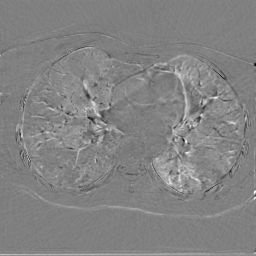} &
\includegraphics[height=3.9cm,width=3.9cm]{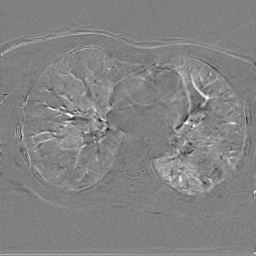} \\ 
\includegraphics[height=3.9cm,width=3.9cm]{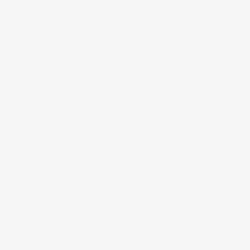} &
\includegraphics[height=3.9cm,width=3.9cm]{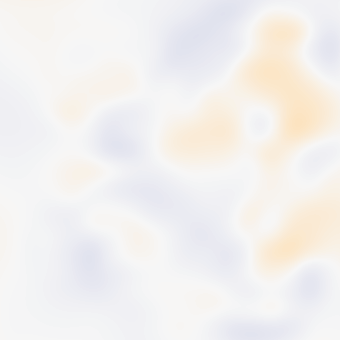} &
\includegraphics[height=3.9cm,width=3.9cm]{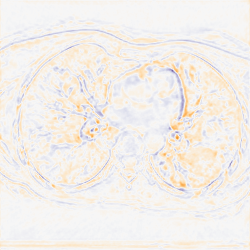} &
\includegraphics[height=3.9cm,width=3.9cm]{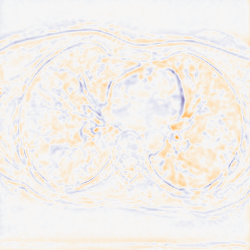} &
\includegraphics[height=3.9cm,width=3.9cm]{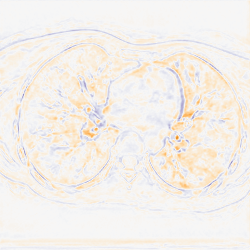} &
\includegraphics[height=3.9cm,width=3.9cm]{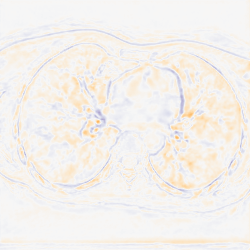} &
\includegraphics[height=3.9cm,width=3.9cm]{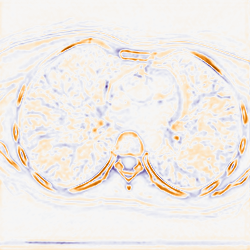} &
\includegraphics[height=3.9cm,width=3.9cm]{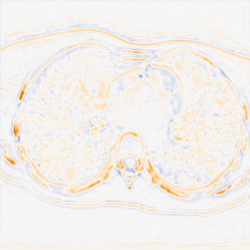} \\ 
\end{tabular}
}
\medskip
\caption{Qualitative evaluation of first patient of \textit{DIR-Lab 4DCT} dataset after applying different registration models. \textsc{Top:} Axial slice of average timepoint after subtracting the first timepoint. \textsc{Bottom:} Axial slice of the Jacobian determinant field averaged over all timepoints.}
\label{fig:qualitative1}
\end{figure}

\subsubsection{Breast MR Dataset}
The \textit{Breast MR} dataset comprises 270 training and 68 evaluation dynamic contrast enhancement series of subjects with extremely dense breast tissue (Volpara Density Grade 4). Each series contains DCE-MRI images ($384 \times 384 \times 60$ voxels with spacing $0.966 \times 0.966 \times 3.000$ mm resampled to $1\text{mm}^3$) acquired on a 3.0T Achieva or Ingenia Philips system at $K=6$ timepoints in the axial plane with bilateral anatomic coverage. The interval between timepoints between the first two timepoints is 90 seconds in which a 0.1 mL dose of gadobutrol per kilogram of body weight is injected at a rate of 1 mL/sec and 1 minute for the other timepoints.

\subsubsection{DIR-Lab 4DCT}
This set contains 4D chest CT scans of 10 different patients that have been acquired as part of the radiotherapy planning process for the treatment of thoracic malignancies. For each patient 3D CT volumes have been acquired at $K=6$ timepoints during the expiratory phase with an average voxel spacing of $1.04 \times 1.04 \times 2.5$. A detailed description of the data acquisition can be found in \cite{castillo2009framework} and \cite{castillo2009four}. The dataset is a subset of the publicly available \textit{DIR-Lab 4DCT} dataset \cite{castillo2009framework} and contains 75 landmark reference points for all timepoints. The volumes have been cropped in-plane to $256 \times 256$ voxels in such a way that all reference points are present. We report performance on the first scan (`Patient 1') and used the others for training.

\section{Experiments and Results}
Using the proposed GroupMorph model, we have performed intra-patient registration of multiple 3D scans acquired at different timepoints in two different ways. First, we performed \textit{all-to-one} registration where the scan acquired at the first time point was used as fixed image and all other scans of the same patient as moving images. Thereafter, we performed \textit{all-moving} registration, where we compute deformation fields for all images and the register to a geodesic average.

To compare performance of the proposed GroupMorph model with state-of-the-art conventional and neural-based methods, we performed registration using Elastix \cite{klein2009elastix} and VoxelMorph \cite{balakrishnan2019voxelmorph,dalca2019unsupervised} .
For registration with Elastix we use  configuration described for registration of DCE-MRI descriped in \cite{gubern2015automated}.
For fair comparison with VoxelMorph, we use the same loss, network architecture and hyper-parameters in VoxelMorph and GroupMorph. We only vary the channels in the first and last layers to match required dimensions and set $\lambda_v=0$ for VoxelMorph.

Experiments were performed using both breast MRI and chest CT scans. For breast MR scans no landmarks were available, hence evaluation was performed with root mean square error (RMSE) and NMI. For chest CT, a set of 75 landmarks was provided at all timepoints within DIR-Lab, hence performance was evaluated by measuring the average distance between the landmarks after registration and over all timepoint combinations. Moreover, for both sets we report percentages of negative Jacobian determinants ($\%\ |J_\phi|<0$) that indicate undesirable folding. Additionally, we evaluate the speed of registrations. 
Results are listed in Table~\ref{tab:quantitative1} \footnote{We measured time to calculate all required deformation fields for Elastix on two 10-core Intel Xeon Silver 4114 2.20GHz CPUs and of the neural-based methods on a NVIDIA GeForce RTX 2080 GPU.}.
They show that the group-wise models perform better in terms of RMSE and NMI on both sets, especially in the \textit{all-moving} case. We find that diffeomorphic registration models (indicated with `-diff') lead to lower amount of negative Jacobian determinants, which means that less folding occurs and indicates that the fields are smoother than those of non-diffeomorphic counterpart.
Average registration times per series (Table~\ref{tab:quantitative1}) show that registration times of the GroupMorph model are significantly lower than the registration times of the baselines, especially compared with Elastix.

Figure \ref{fig:qualitative1} illustrates the axial slices of the first volume in the \textit{DIR-Lab} dataset. Subtraction images illustrate improved alignment for all registration models, especially at borders and in particular for \textit{all-moving} group-wise models. From the Jacobian determinant fields, we can observe particularly strong local expansions and compression in the \textit{all-moving} models and most notably in the non-diffeomorphic variant.

\newpage
\section{Discussion and Conclusion}

\begin{table}[t]
\centering
\caption{Quantitative performance using different registration methods. \textsc{Left}: Models and average registration time per serie. \textsc{Center}: RMSE and NMI on \textit{Breast MR}. \textsc{Right:} RMSE, NMI and average displacment on \textit{DIR-Lab 4DCT} evaluation set.}
\label{tab:quantitative1}
\resizebox{\linewidth}{!}{
\begin{tabular}{l c c c c l c c c c }
\hline
\multicolumn{1}{c}{\multirow{2}{*}{\pbox{2cm}{Model}}} &
\multicolumn{1}{c}{\multirow{2}{*}{\pbox{2cm}{Time (s) } }} &
\multicolumn{3}{c}{\small \textit{Breast MR}} &
\multicolumn{4}{c}{\small \textit{DIR-Lab 4DCT}}
\\
\cline{3-5} \cline{7-10}
& - & RMSE & NMI & $\%\ |J_\phi|<0$ &
& RMSE & NMI & Avg Displacement ($mm$) & $\%\ |J_\phi|<0$ \\
\hline 
No Registration (Identity) & -
& 24.15 & 0.24 & - &
& 63.39 & 0.47 & 1.83 $\pm$ 2.24 & - \\
Elastix & 652
& 23.42 & 0.25 & 0 &
& 36.84 & 0.49 & 1.09 $\pm$ 1.30 & 0 \\
\hline
VoxelMorph & 5 
& 22.43 & 0.26 & 3.31e-2 & 
& 28.42 & 0.50 & 1.18 $\pm$ 1.33 & 4.32e-6 \\
VoxelMorph-diff & 5
& 22.12 & 0.26 & 1.55e-4 &
& 25.37 & 0.51 & 1.12 $\pm$ 1.33 & 7.79e-7 \\
\hline
\textit{All-to-one} GroupMorph \textbf{\small (ours)} & \textbf{2}
& 21.97 & 0.28 & 4.20e-5 &
& 27.11 & 0.51 & 1.16 $\pm$ 1.36 & 3.05e-5 \\
\textit{All-to-one} GroupMorph-diff \textbf{\small (ours)} & \textbf{2}
& 22.00 & 0.27 & 1.60e-5 & 
& 25.26 & 0.51 & \textbf{1.08 $\pm$ 1.35} & 2.60e-7 \\
\textit{All-moving} GroupMorph \textbf{\small (ours)} & \textbf{2}
& \textbf{21.47} & \textbf{0.32} & 2.00e-4 &
& 23.15 & 0.53 & 1.39 $\pm$ 1.32 & 4.98e-6 \\
\textit{All-moving} GroupMorph-diff \textbf{\small (ours)} & \textbf{2}
& 21.57 & \textbf{0.32} & 1.90e-4 &
& \textbf{22.09} & \textbf{0.54} & 1.52 $\pm$ 1.34 & 9.74e-7 \\
\hline
\end{tabular}
}
\end{table}

In this study, we proposed a variational and diffeomorphic learning-based registration method for application in group-wise image registration. We evaluated our approach using intra-patient registration with sets of multiple 3D scans of two different modalities (MR and CT) showing different anatomies (breast and chest). In addition, we provided a likelihood based on normalized mutual information, a well-performing image similarity metric in registration, between multiple images and a prior that allows for explicit control over the viscous fluid energy to effectively control regularization of deformation fields.

In spite of using diffeomorphic transformations, we observed some negative Jacobian determinants likely resulting from approximation errors. Hence, future work could investigate options for further improvements. It can be expected that training with more data, further tuning of the network architecture and hyper-parameters will lead to higher performance. Other potential advancements include a 4D network \cite{choy20194d} to more explicitly model the temporal component, fine tuning of $\z$ after inference (see dotted grey arrow in Figure 2), time-dependent velocity fields to allow more flexible diffeomorphic transformations, and a bi-directional cost \cite{mghimproved} and non-diagonal covariance \cite{dalca2019unsupervised} to further improve smoothness of the deformation fields.

To conclude, we found that our GroupMorph model can simultaneously register multiple scans with performance similar to conventional registration with Elastix and learning-based VoxelMorph, enabling close to real-time group-wise registration without delaying clinical reading.

\section*{Acknowledgements}
Authors would like to thank Jorrit Glastra, Tim Henke and Rob Hesselink for insightful discussions and support.

\newpage
\bibliographystyle{splncs04}
\bibliography{biblio}

\newpage
\section*{Supplementary Material}

\subsection*{Derivation of Main Loss}
By following \cite{kingma2013auto} and the Supplementary Material of \cite{dalca2019unsupervised} adjusting for our proposed likelihood and prior we obtain

\begin{align*}
&\mathcal{L}(\theta;\x,\y) =\\
&= -\E_q[\log p(\x|\y,\z) + \text{KL}[q_\theta(\z|\x,\y)||p(\z)] \\
&= -\E_q\left[\log \xoverbrace{\exp \frac{1}{K(K-1)}\sum_{i\neq j} \text{NMI}(\phi_{\z_i} \circ \x_i, \phi_{\z_j} \circ \x_j)}^{ p(\x|\y,\z)}\right]
+ \text{KL}\left[\mathcal{N}(\z;\mmu_\z,\SSigma_\z)||\mathcal{N}(\z;0,\LLambda)\right] + \text{const.}\\
&= \frac{1}{K(K-1)}\sum_{i\neq j} \E_q\left[
 -\text{NMI}(\phi_{\z_i} \circ \x_i, \phi_{\z_j} \circ \x_j)\right]
+ \frac{1}{2}\left[
\log \frac{|\LLambda_\z^{-1}|}{|\SSigma_\z|} - 3d + \text{tr}(\LLambda_\z \SSigma_\z) + \mmu_\z^T \LLambda \mmu_\z]
\right] + \text{const.}\\
\end{align*}

\noindent
We know that $\log |\LLambda|$ is constant, $\log|\SSigma_\z|=\text{tr}\log \SSigma_\z$, and $\text{tr}(\LLambda_\z \SSigma_\z)=\text{tr}((\lambda_u \mathbf{D} - \mathbf{A} + \lambda_v\mathbf{I})\SSigma_\z))=\text{tr}((\lambda_u \mathbf{D} + \lambda_v \mathbf{I})\SSigma_\z)$, and approximating the expectation with 1 sample $z_k \sim q_z$, to find

\begin{align*}
&\mathcal{L}(\theta;\x,\y) = 
\frac{1}{K(K-1)}\sum_{i\neq j} -\text{NMI}(\phi_{\z_i} \circ \x_i, \phi_{\z_j} \circ \x_j)
+ \frac{1}{2} \Big[ \text{tr}((\lambda_u \mathbf{D} + \lambda_v)\SSigma_\z - \log \SSigma_\z) + \mmu_\z^T \LLambda^{-1} \mmu_\z \Big] + \text{const.}
\end{align*}

\end{document}